\documentclass[12pt]{iopart}

\usepackage{epsfig}  
\usepackage{lineno}

\begin{document}

\title[Conditions for ELM suppression in ASDEX Upgrade]{Experimental conditions to suppress edge localised modes by magnetic perturbations in the ASDEX Upgrade tokamak}

\author{W~Suttrop$^1$, A~Kirk$^2$, V~Bobkov$^1$, M~Cavedon$^1$, M~Dunne$^1$, R~M~McDermott$^1$,
  H~Meyer$^2$, R~Nazikian$^3$, C~Paz-Soldan$^4$, D~A~Ryan$^2$, E~Viezzer$^{1,5}$, M~Willensdorfer$^1$,
  the ASDEX Upgrade\footnote{See A~Kallenbach et al, Nucl Fus {\bf 57} (2017) 102015}
  and MST1\footnote{See H~Meyer et al, Nucl Fus {\bf 57} (2017) 102014} Teams}

\address{$^1$Max Planck Institute for Plasma Physics, Boltzmannstrasse 2, 85748 Garching, Germany}
\address{$^2$CCFE Culham Science Centre, Abingdon, Oxon, OX14 3DB, UK}
\address{$^3$Princeton Plasma Physics Laboratory, PO Box 451, Princeton, New Jersey 08543-0451, USA}
\address{$^4$General Atomics, PO Box 85608, San Diego, California 92186-5608, USA}
\address{$^5$Dept. of Atomic, Molecular, and Nuclear Physics, University of Seville, Avda. Reina Mercedes, 41012 Seville, Spain}

\ead{suttrop@ipp.mpg.de}
\vspace{10pt}
\begin{indented}
\item[]Draft, \today
\end{indented}

\begin{abstract}
  Access conditions for full suppression of Edge Localised Modes (ELMs)
  by Magnetic Perturbations (MP) in low density high confinement mode (H-mode) plasmas
  are studied in the ASDEX Upgrade tokamak.
  The main empirical requirements for full ELM suppression in our experiments are:
  1. The poloidal spectrum of the MP must be aligned for best plasma response
  from weakly stable kink-modes, which amplify the perturbation,
  2. The plasma edge density must be below a critical value, $3.3 \times 10^{19}$~m$^{-3}$.
  The edge collisionality is in the range
  $\nu^*_i = 0.15-0.42$ (ions) and $\nu^*_e = 0.15-0.25$ (electrons).
  However, our data does not show that the edge collisionality is the critical
  parameter that governs access to ELM suppression.
  3. The pedestal pressure must be kept sufficiently low to avoid destabilisation of
  small ELMs. This requirement implies a systematic reduction of pedestal pressure
  of typically 30\% compared to unmitigated ELMy \mbox{H-mode} in otherwise similar plasmas.
  4. The edge safety factor $q_{95}$ lies within a certain window.
  Within the range probed so far,
  $q_{95}=3.5-4.2$, one such window, $q_{95}=3.57-3.95$ has been identified.
  Within the range of plasma rotation encountered so far, no apparent threshold of
  plasma rotation for ELM suppression is found. This includes cases with large
  cross field electron flow in the entire pedestal region.
\end{abstract}

%
\vspace{2pc}
\maketitle
%
%


\section{Introduction}

The transient heat load onto the first wall associated with
the edge localised mode (ELM) instability
is a main concern for the next step fusion device, ITER, 
and for a fusion reactor.
Complete ELM suppression by small magnetic perturbations (MP)
to the axisymmetric tokamak,
first demonstrated in DIII-D \cite{EVANS04A},
is one of the main methods considered for ITER
to ensure an appropriate first wall lifetime and to
prevent an excessive contamination of the plasma with heavy impurities
produced by ELM-induced wall erosion \cite{LOARTE_NF54_2014_033007}
while maintaining the favourable properties
of high confinement mode (H-mode).
ELM suppression has been reproduced recently in 
KSTAR \cite{JEON_PRL109_2012_035004} and EAST \cite{SUN_PRL117_2016_115001},
albeit at higher edge pedestal collisionality than in DIII-D and ITER.

ASDEX Upgrade (AUG) is equipped with two rows of MP coils,
each with eight toroidally distributed
in-vessel saddle coils \cite{SUTTROP09A}.
They are capable of producing a peak MP field,
measured at the plasma surface, of the order of $10^{-3} B_t$, where $B_t \leq 3.2$~T
is the toroidal magnetic field in AUG.
Independent MP coils power supplies for each MP coil \cite{TESCHKE_FED96_2015_171}
allow us to vary the poloidal structure of the MP field within a plasma discharge.
This flexibility allows us to rotate MP fields
with toroidal mode number $n=1-3$
rigidly for measurements of the plasma response
\cite{WILLENSDORFER_PPCF58_2016_114004,WILLENSDORFER_NF57_2017_116047}
and to vary the phase between the upper and lower coil ring
(dubbed the ``differential phase'') in order to
vary the relative strength of resonant and non-resonant
spectral modes \cite{SUTTROP_FED88_2013_446}.

\begin{figure}[t]
  \centering
  \includegraphics[width=0.5\columnwidth]{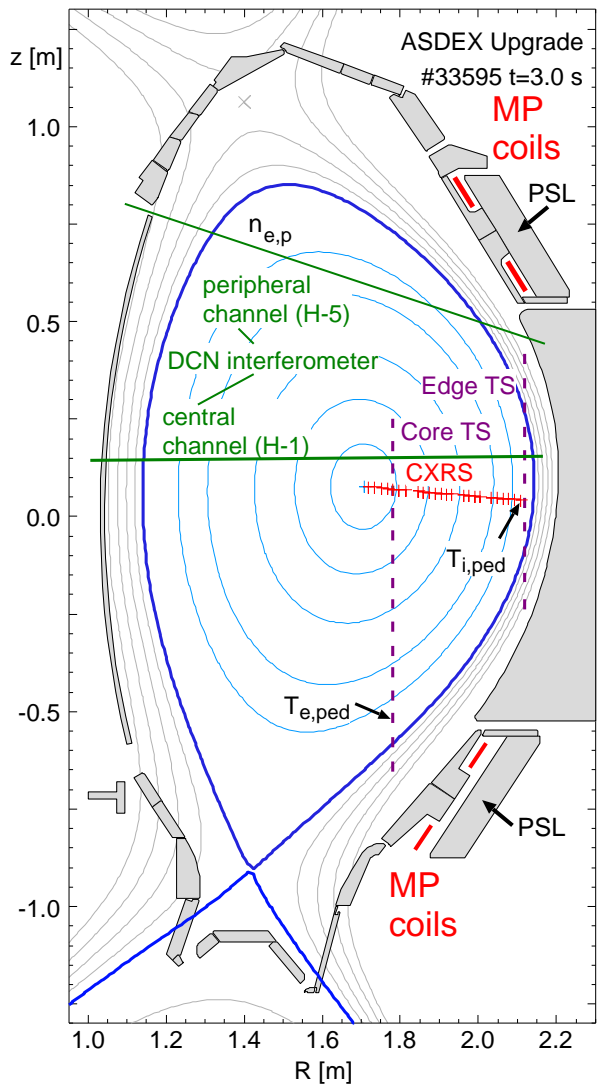}

  \caption{Cross section of the ELM suppression plasmas studied, with MP coils positions,
           and sightlines of some of the main diagnostics overlayed (see text).}
  \label{fig:xsect}
\end{figure}

With $n=1, 2$ and $4$ magnetic perturbations, a significant reduction
of the energy losses associated with
individual ELMs (ELM mitigation) has been obtained at
high \cite{SUTTROP11A} and low pedestal collisionality \cite{KIRK_NF55_2015_043011}.
Attempts to fully suppress ELMs in stationary H-mode plasmas
in AUG had long been unsuccessful.
In a recent matching experiment of AUG and DIII-D \cite{NAZIKIAN_FEC2016},
the plasma shape has been identified as a critical parameter.
In plasmas with elevated upper triangularity, complete suppression of ELMs by
magnetic perturbations has been observed for the first time
in AUG \cite{NAZIKIAN_FEC2016, SUTTROP_PPCF59_014049}.
The decisive influence of plasma shaping has been attributed
to higher pedestal pressure at elevated triangularity and hence,
stronger amplification of the external MP by plasma response \cite{NAZIKIAN_FEC2016}.
Apart from plasma shape, other experimental conditions appear to be crucial
for attaining full suppression of ELMs.
The initial success of ELM suppression in AUG enabled a recent study of
access parameters, which is reported in this paper.

The paper is organised as follows:
The experimental setup used to suppress ELMs in H-mode plasmas is described in
section \ref{sec:Scenario}.
The role of several parameters for accessing ELM suppression is
studied in section \ref{sec:Conditions}, namely the 
resonant alignment of the MP, the choice of edge safety factor,
plasma edge density and collisionality, and the role of plasma rotation.
Finally (section \ref{sec:Discussion}), we
discuss the implications of our results for ELM suppression models.


\section{ELM suppression by magnetic perturbations}
\label{sec:Scenario}

For the present experiment, the ELM suppression scenario described
in Ref. \cite{SUTTROP_PPCF59_014049}
is used throughout, in particular the nominal plasma shape.
Fig. \ref{fig:xsect} shows the cross section of a typical plasma, with
the poloidal contours of in-vessel structures, MP coils and selected
diagnostics sightlines.
The two rows of MP saddle coils in AUG are located at the low field side,
above and below midplane.
They are mounted onto two massive copper conductors wired as
an $n=0$ saddle loop, termed the Passive Stabilising Loop (PSL).
The PSL serves to reduce the vertical growth rate of the elongated AUG
plasma by induction of a radial field that counter-acts
vertical plasma position excursions.
Some of our experiments (see section \ref{sec:PlasmaResponse})
employ fast transients of MP coil currents.
These transients induce eddy currents in the PSL conductor behind each
individual MP coil that decay resistively and cause the evolution of the
total vacuum field (from PSL plus MP coil) to lag behind the MP coil current.
These eddy currents can significantly affect the amplitude and phase of the
magnetic perturbation, as seen in section \ref{sec:PlasmaResponse}, and must
be taken into account during MP coil transients.
The total vacuum field including PSL response is calculated by a magnetodynamic
finite element model as a function of frequency, from which a continuous
complex transfer function is obtained \cite{SUTTROP09B}.
Because of the proximity of the MP coil conductors and the PSL, compared to
the distance to the plasma surface, we can express the shielding effect of the
PSL as a lumped, effective coil current for which the vacuum field is calculated.

In the present study, we use full profiles with sufficient resolution in the
H-mode pedestal region to represent gradients in the edge transport barrier
- sightlines of some measurements are shown in Fig. \ref{fig:xsect}.
This includes edge and core Thomson scattering
(electron density, $n_e$ and electron temperature $T_e$)
using two different vertical laser beam lines and horizontal observation,
and charge exchange recombination spectroscopy of boron ($B^{5+}$) for ion
temperature ($T_i$) and impurity toroidal rotation (v$^{B5+}_\mathrm{tor}$).
Continuous time traces of edge and core electron density,
electron temperature, ion temperature and toroidal impurity rotation
are taken from a peripheral and a central DCN (deuterated cyanide)
interferometer channel (H-5 and H-1 chords, respectively),
core Thomson scattering observation channel 14 and
core CXRS observation channel 24, as indicated in the figure.
The edge interferometer H-5 chord is tangential at $\rho_p = 0.84$ for this
plasma shape and position, which is representative for the pedestal density in
our discharges. Below, this measurement is denoted as $n_{e,p}$.

\begin{figure}[t]
  \centering
  \includegraphics[width=0.7\columnwidth]{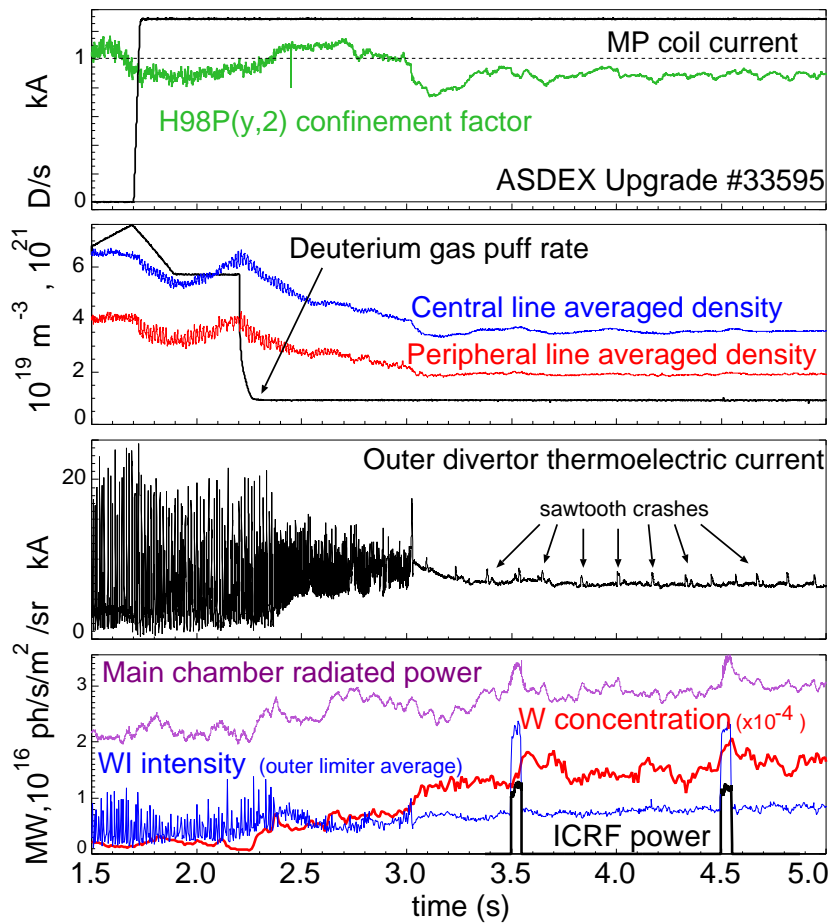}

  \caption{Time traces of ASDEX Upgrade discharge 33595 showing ELM suppression
    after $t=3.0$~s. ICRF pulses at $t=3.5$~s and $t=4.5$~s in
    monopole phasing provoke
    increased tungsten influx from the outer limiters -- the plasma
    tungsten concentration recovers quickly.}
  \label{fig:Traces_AUG_33595}
\end{figure}

ELM suppression discharges are performed after boronisation of the vacuum vessel wall
in order to obtain the lowest possible plasma density in H-mode.
Time traces of discharge $33595$ are shown in 
Fig.~\ref{fig:Traces_AUG_33595} as an example of long stationary ELM suppression.
The startup is similar to conventional H-mode plasmas. However the MP coils
are switched on at an early time ($t=1.7$~s) in H-mode in order to reduce
the ELM size. At $t=2.2$~s the gas puff rate is reduced to a very low level,
$1 \times 10^{21}$~D/s, which leads to a phase with
increased ELM frequency and reduced ELM losses $t=2.35 - 3$~s, during which
the central and peripheral plasma densities continuously decrease.
This ``pump-out'' phenomenon due to the application of MP in low density plasmas
is commonly observed in AUG and other
experiments \cite{LEUTHOLD_PPCF59_2017_55004}.
At $t=3$~s, ELM activity stops completely for the remainder of the H-mode flat top.
The H-mode confinement factor $H98P_{y,2}$ \cite{IPB_NF39_1999_2175}
in the initial ELMy phase is $H98P_{y,2}=1.0$ and drops
to $H98P_{y,2}=0.9 - 0.95$ at later times during the suppressed phase.
Full suppression of ELMs is indicated by a large number of signals, e.g. the
outer divertor thermoelectric current (third panel),
which is a reliable indicator of divertor
temperature and, therefore, ELM-related heat pulses.
In the suppression phase, transient heat pulses from sawtooth crashes are observed;
however, the magnetic measurements indicate that in most cases they do not trigger ELMs.
It should be noted that in reference discharges without MP but otherwise
identical plasma shape and actuator trajectories, the ELM frequency decreases
and plasma density remains high after the gas puff is reduced. 

As a special feature, AUG has a fully tungsten-clad first wall \cite{NEU10A}.
Stable H-mode operation with a metal wall requires net outward
transport of heavy impurities to avoid radiative collapse of the
plasma core \cite{NEU10A}, which is normally assisted by gas puffing
in order to avoid density profile peaking and to ensure a sufficiently large
ELM frequency.
In AUG, ELM suppression can only be achieved without strong gas puff.
Therefore, it is important to verify that impurity accumulation can be
avoided in the absence of ELMs.
Short pulses of power in the Ion Cyclotron Range of Frequencies 
(ICRF, see fourth panel of Fig. \ref{fig:Traces_AUG_33595}) are applied
to inject tungsten impurities into discharge $33595$.
Instead of using an optimum phase and power distribution between the straps of 
the newly installed 3-strap ICRF antennas \cite{BOBKOV_NF2016_56_84001}
in order to minimise the induced radiofrequency (RF) currents in the antenna box
and the associated RF sheaths,
we deliberately apply the same phase to all of the three antenna straps
(monopole phasing) to enhance the RF sheaths
and to sputter tungsten from the antenna limiters.
The resulting tungsten influx can be seen
as an increased intensity of WI (neutral tungsten) spectroscopic lines.
A small increase of tungsten concentration
(higher charge states measured by an X-ray spectrometer)
and main chamber radiated power follows and recovers
to a steady state after about 200~ms, with a time constant
slightly above the energy confinement time, $\tau_W \approx 1.2 \tau_E$.
Hence, a particle transport mechanism is active which is not only causing the
``pump-out'' of main ions, but also flushes heavy impurities.
This is consistent with the observation of outward transport of
medium-Z impurities (fluorine) in DIII-D \cite{GRIERSON_POP22_2015_55901}.


\section{Access conditions to ELM suppression}
\label{sec:Conditions}


%
%
%


\subsection{Resonant magnetic perturbation}
\label{sec:PlasmaResponse}

\begin{figure}[t]
  \centering
  \includegraphics[width=0.7\columnwidth]{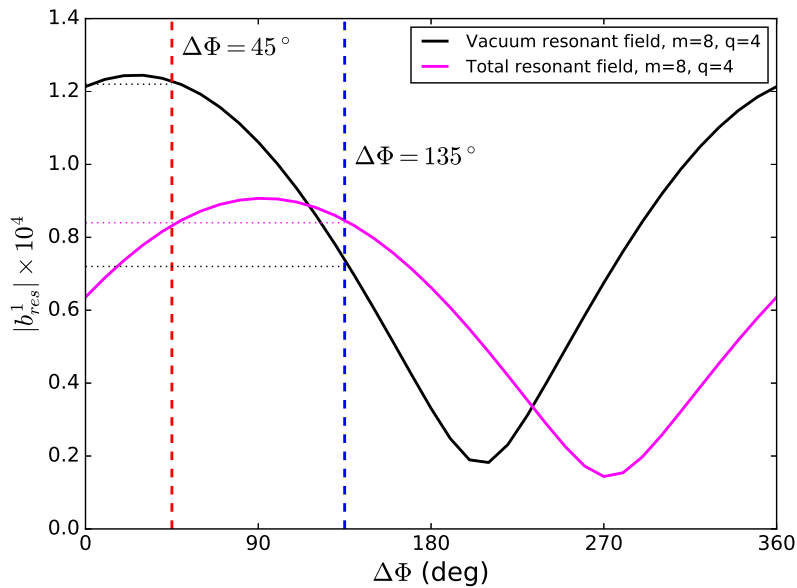}

  \caption{Predicted $m=8$, $n=2$ resonant magnetic perturbation field at the $q=4$ surface
    for fixed MP coil current amplitude, $I_\mathrm{MP}=1.7$~kA,
    normalised to the equilibrium magnetic field,
    as a function of differential phase angle $\Delta \Phi$ between
    upper and lower MP coil current patterns for pure vacuum response (black) and
    including the plasma response, as calculated by the MARS-F model (magenta).
    Experimental test cases are marked by dashed vertical lines.}
  \label{fig:MARS-F_deltaphi}
\end{figure}

The relevance of the poloidal MP spectrum for access to ELM suppression can be tested
by varying the relative phase of poloidally separated, toroidally equidistantly spaced
MP coil sets, as has been done before
using the two rows of 6 in-vessel saddle coil (I-coils) for $n=2$ perturbations
in DIII-D \cite{LANCTOT_NF53_2013_83019}.
The finite number of MP coils in the toroidal direction
($n_\mathrm{coils}=8$) leads to spatial
aliasing, i.e. leakage of the applied $n=2$ MP pattern to
$n_\mathrm{alias} = n_\mathrm{coils} - n = 6$.
Apart from the $n=6$ sideband, the aliasing effect in AUG is a small
modulation of the $n=2$ amplitude as the differential phase is varied.

\begin{figure*}[t]
  \centering
  \includegraphics[width=0.9\textwidth]{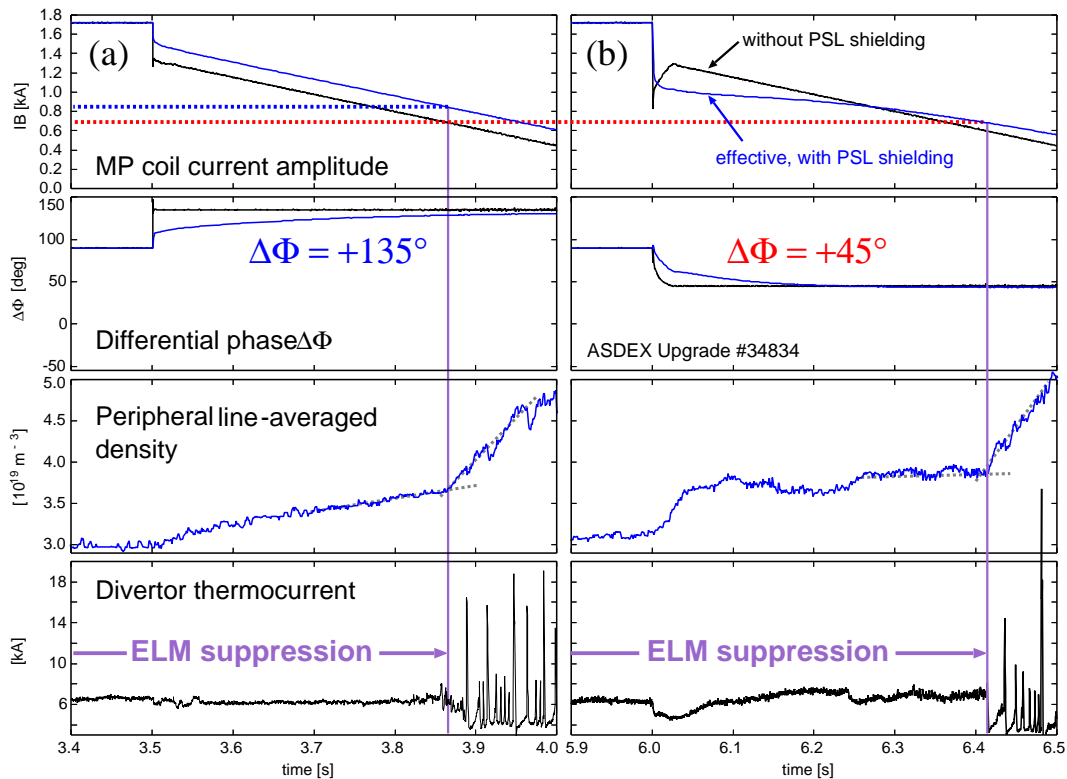}

  \caption{Measurements of the $I_{MP}$ threshold for back transition to ELMy H-mode
    for two different values of the phase difference $\Delta \Phi$
    of upper and lower MP coil rings. The MP coil current amplitude (top panel)
    and differential phase (second panel) are corrected for vacuum field shielding
    by the neighbouring PSL conductor (blue curves).}
  \label{fig:IMPthresh}
\end{figure*}

The effect of differential phase variation on the calculated
resonant magnetic perturbation is
demonstrated in Fig. \ref{fig:MARS-F_deltaphi}.
The $n=2$, $m=8$ resonant radial magnetic field amplitude $b_\mathrm {1,res}$
at the $q=4$ surface for fixed MP coil current amplitude, $I_\mathrm{MP}=1.7$~kA,
normalised to the total magnetic field is shown
as a function of the differential phase $\Delta \Phi$
(defined in Ref. \cite{SUTTROP_PPCF53_2011_124014}).
Two figures of merit are considered: a pure vacuum response
(no helical plasma currents induced by the applied MP, black curve),
and the resonant field including the plasma response,
which is calculated using the
linear resistive MARS-F fluid model \cite{RYAN_PPCF57_2015_95008} (magenta curve).
The underlying MHD equilibrium is that of the ELM suppression scenario
described in section \ref{sec:Scenario}.
The maximum vacuum response ($\Delta \Phi \approx 30^\circ$) corresponds to
alignment of the MP coil phasing with the plasma magnetic field.
The plasma response to the vacuum field is two-fold in nature. Firstly,
the resistive response to field-aligned MP is partially shielded
by helical currents on resonant rational
surfaces which are driven by flows perpendicular to the magnetic field \cite{BECOULET12A}.
Secondly, the MP is amplified by marginally stable ideal MHD modes, driven by the edge
pressure gradient and edge current (which is dominated by the bootstrap current in
the H-mode edge gradient region) \cite{LANCTOT_NF53_2013_83019}.
Because of poloidal mode coupling due to toroidicity and vertical elongation of the torus,
these modes produce a resonant response \cite{RYAN_PPCF57_2015_95008}.
This can be seen in Fig. \ref{fig:MARS-F_deltaphi}
particularly for $\Delta \Phi = 120^\circ -  250^\circ$, where the
plasma-driven resonant response exceeds the unshielded vacuum
response.
It should be noted that while the magnitude of the plasma response depends sensitively
on the pressure and edge current density profiles, the differential phase for optimum
plasma response depends weakly on plasma pressure, as found in MARS-F calculations
for a scan of $\beta_N$ using the ASDEX Upgrade arrangement of MP coils and
an ASDEX Upgrade equilibrium as the base case \cite{Ryan_PPCF59_2017_024005}.

We consider two different cases for an experiment that highlights the importance
of the plasma response: 
(a) $\Delta \Phi=+135^\circ$ and (b) $\Delta \Phi=+45^\circ$, 
in which the calculated resonant vacuum field differs by about a factor of two (for the
same effective MP coil current),
while the MP field including plasma response is similar (see Fig. \ref{fig:MARS-F_deltaphi}).
If we assume that maintaining ELM suppression requires that
the resonant field $b^1_\mathrm{res}$ remains above a certain fixed threshold value,
then the measured effective MP coil current thresholds for the two values of
$\Delta \Phi$ should be inversely proportional to one of the two
calculated response curves in Fig. \ref{fig:MARS-F_deltaphi}.
Since the ratio of the response fields for these $\Delta \Phi$ values differ
significantly for the vacuum-only and total resonant response models,
our experiment can discriminate between the two models.
Fig. \ref{fig:IMPthresh} shows time traces of the two cases,
which are examined in different time intervals in discharge 34834.
In each case, reproducible initial conditions are set by a preceding phase with
optimum plasma response $\Delta \Phi = 90^\circ$ and maximum MP coil current.
This results in an initially stationary ELM suppression phase with low
plasma density, $n_e = 3.0 \times 10^{19}$~m$^{-3}$, in order to obtain a similar
plasma response in both cases.
The MP coil current phasing is then switched to the $\Delta \Phi$ value for the
respective case and the MP coil current amplitude is slowly ramped down to measure
the threshold for losing ELM suppression.
The upper two panels of Fig. \ref{fig:IMPthresh} show the $n=2$ spatial amplitude and phase,
as obtained from actual MP coil currents (black time traces) and derived
from effective MP coil currents that take into account the shielding by the PSL
(blue time traces). One can clearly see that the presence of the PSL currents affects
both amplitude and differential phase of the MP field, therefore the
effective MP coil current must be used for this comparison.

Loss of ELM suppression is detected by a reversal to a classical ELM-free phase,
characterised by a rapid increase of plasma density, followed by large ELM activity.
Just before ELM suppression is lost
(at the times denoted by vertical magenta lines), the plasma density
(third panel in Fig. \ref{fig:IMPthresh}) has increased compared
to the begin of the coil current ramp, but is similar in the two cases.
For cases (a) and (b), with similar total (vacuum plus plasma) response,
the effective MP coil current amplitude threshold
is similar, $I_\mathrm{MP} = 820$~A and $700$~A, respectively,
while the resonant (field-aligned) vacuum field (Fig. \ref{fig:MARS-F_deltaphi})
differs by a factor of two.
This comparison shows that the plasma response, i.e. coupling of the applied MP field
to amplifying ideal MHD modes, is essential to maintain ELM suppression.


\subsection{Low edge density and collisionality}
\label{sec:EdgeDensCollisionality}

It can be noted from discharge 33595 (shown in Fig. \ref{fig:Traces_AUG_33595})
that the application of the MP at $t=1.7$~s with correct phasing is a necessary,
but not sufficient condition for ELM suppression, as ELM activity continues until
$t=3.0$~s.
After the gas puff is reduced to a minimum, small ELMs are encountered.
The plasma density in this phase slowly decreases, until the ELM activity ceases.
Therefore, the suppression of ELMs appears to depend on achieving a low plasma density
in H-mode.
Right after the transition to ELM suppression
(at $t=3.0$~s in Fig. \ref{fig:Traces_AUG_33595}) the density drops further and then
levels at a stationary low value for the entire ELM suppression time interval.
Hence, the outward particle transport induced by the MP (the ``pump-out'') increases
during ELM suppression compared to the previous ELM mitigation phase.

\begin{figure*}[t]
  \centering
  \includegraphics[width=0.49\textwidth]{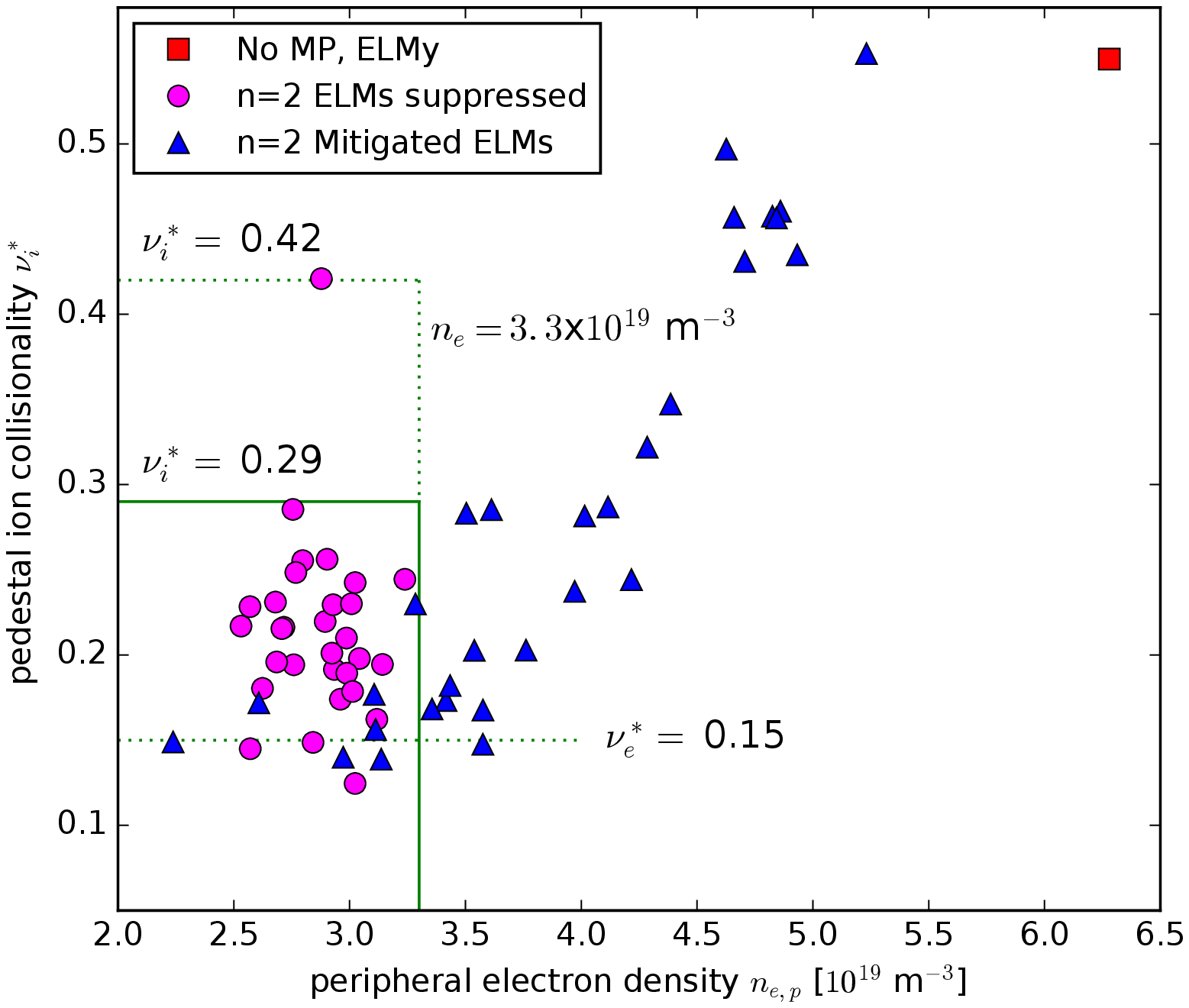}
  \includegraphics[width=0.49\textwidth]{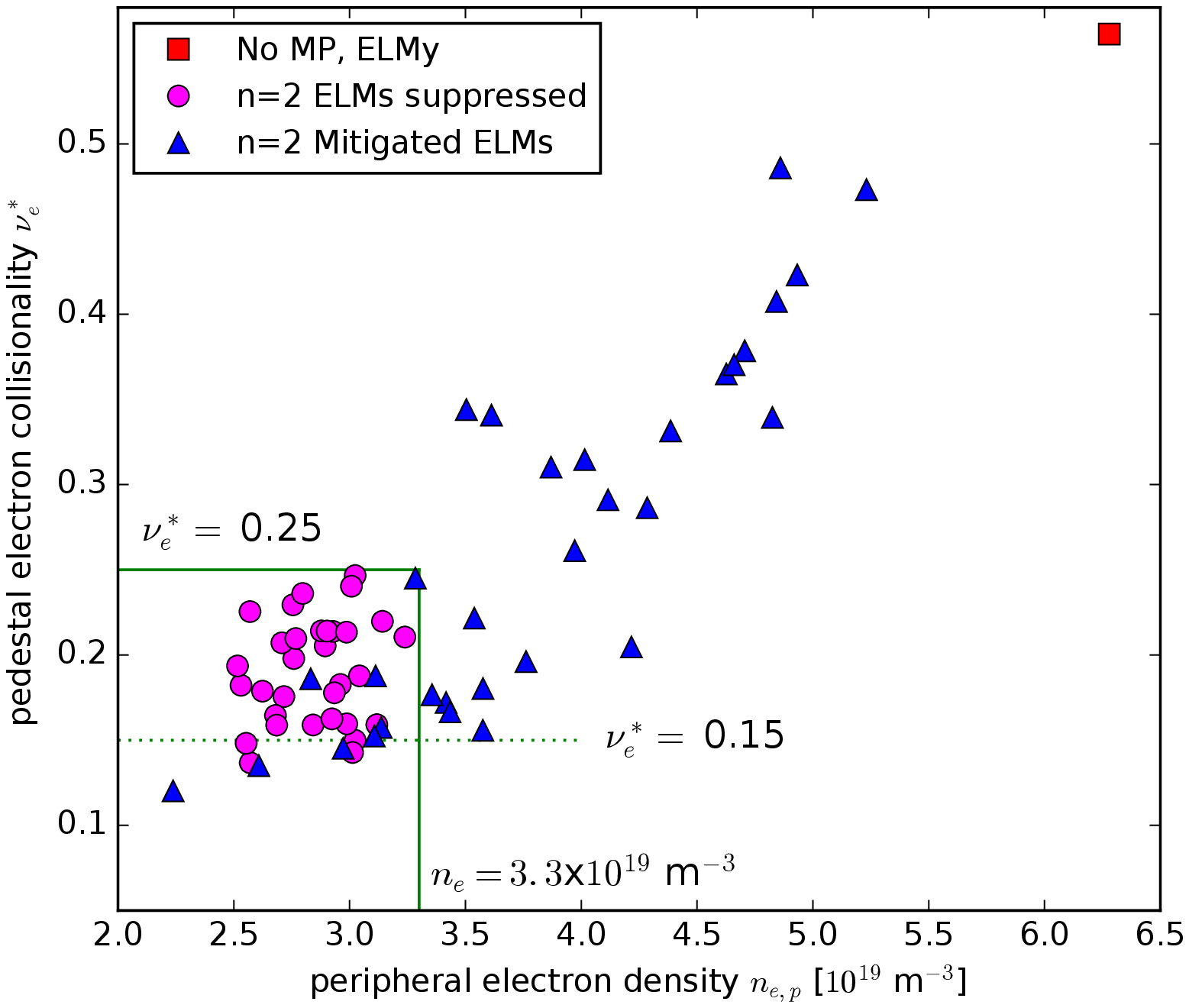}

  \caption{Pedestal collisionality of ions (left) and electrons (right)
    vs. peripheral electron density for phases
    (duration $\Delta t>50$~ms) with ELM suppression (magenta),
    large and small ELMs while $n=2$ MP is applied and unmitigated ELMs without MP.
    Bounding values of $\nu^*_i$, $\nu^*_e$ and $n_{e,p}$, drawn as solid
    and dotted lines, are discussed in the text.}
  \label{fig:nelp_nuped}
\end{figure*}

In an attempt to identify the physically relevant edge parameter for access to
ELM suppression, we can examine the data base of ELM suppression experiments carried out
in AUG so far. This comprises a total of 191 time slices from 44 discharges which
all have the same nominal plasma shape and $B_t = -1.8$~T.
The plasma current is varied between $I_p = 0.7$ and $1.0$~MA
with $I_p = 0.9$~MA in most cases, and the plasmas
are heated with $4-8$~MW neutral beam injection (NBI) power and up to $2.8$~MW
central third harmonic electron cyclotron resonance heating (ECRH) power.
Fig. \ref{fig:nelp_nuped} shows 
the neoclassical pedestal collisionality of ions (left) and electrons (right),
as defined in Refs. \cite{Sauter_PoP6_1999_2834} (Eq.~18) and
\cite{HAGER_PP23_2016_42503} (Eq.~1), 
plotted against the peripheral line-averaged density $n_\mathrm{e,p}$.
All cases shown in the figure use $\Delta \Phi = 90^\circ$, which corresponds to
optimal MP alignment at $I_p=0.9$~MA.


Three data sets are included: ELM suppression (magenta circles),
mitigated ELMs with $n=2$ MP (blue triangles)
and one reference case (red square) without MP but
same low fuelling rate, showing higher plasma density and unmitigated, large ELMs.
Only time intervals with stationary plasma parameters, averaged over
$100$~ms or longer are considered.
All ELM suppression cases are bounded by
$n_\mathrm{e,p} \leq 3.3 \times 10^{19}$~m$^{-3}$
and $\nu^*_\mathrm{i,ped} \leq 0.42$ and $\nu^*_\mathrm{e,ped} \leq 0.25$.
The variation of $\nu^*_\mathrm{i,ped}$ and $\nu^*_\mathrm{e,ped}$
at fixed $n_\mathrm{e,p}$ is mainly due to
variations of the ion and electron pedestal temperature,
$T_\mathrm{i,ped}$ and $T_\mathrm{e,ped}$, respectively.

\begin{figure}[t]
  \centering
  \includegraphics[width=0.7\columnwidth]{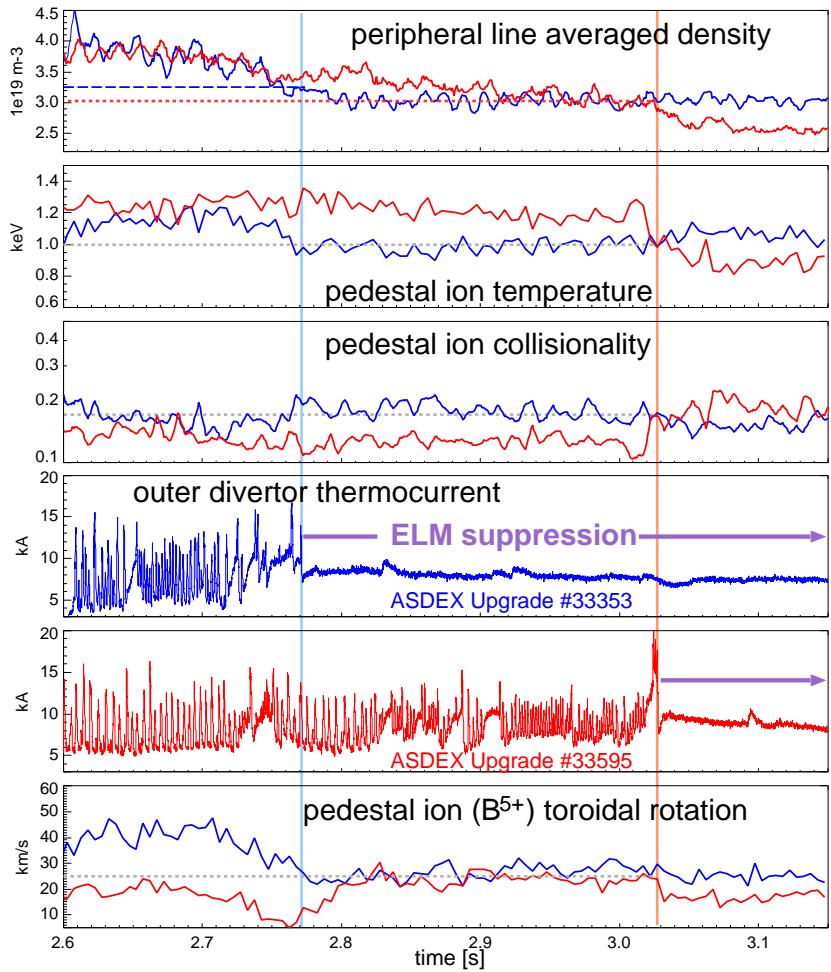}

  \caption{Comparison of pulse $33595$ (red, transition to ELM suppression at $t=3.03$~s)
    with pulse $33353$ (black, transition at $t=2.77$~s).
    The transition in pulse $33595$ is delayed, despite lower collisionality and similar
    plasma rotation as in pulse $33353$.}
  \label{fig:LateTransition_33595}
\end{figure}

Two observations can be made from Fig. \ref{fig:nelp_nuped}.
Firstly, there are no cases with ELMs at $n_\mathrm{e,p} \leq 3.3 \times 10^{19}$~m$^{-3}$
but collisionality larger than those with ELM suppression.
Therefore, we cannot conclude from our data whether there is an upper collisionality limit. 
Secondly, for $n_\mathrm{e,p} \leq 3.3 \times 10^{19}$~m$^{-3}$ small ELM activity 
is still found at low $\nu^*_\mathrm{i,e,ped} \leq 0.15$,
i.e. at high $T_\mathrm{i,ped}$ and high $T_\mathrm{e,ped}$.
This finding points to an upper pedestal temperature limit
for ELM suppression.
We therefore examine in more detail two discharges, 33353 with early ELM suppression
(at $t=2.77$~s) and 33595, where ELM suppression is delayed
to $t=3.028$~s despite reaching low $\nu^*_\mathrm{i,ped}$ early.
Fig. \ref{fig:LateTransition_33595} shows time traces for these two pulses
(33353: blue lines, 33595: red lines).
All plasma control request waveforms for the two shots are identical.
Plasma parameters at the transition to ELM suppression are marked with
dashed lines.
The main difference between the shots is that in 33595,
$T_\mathrm{i,ped} \sim 1.2$~keV in the extended ELMy phase ($t=2.77-3.028$~s),
well above $T_\mathrm{i,ped} \sim 1.0$~keV in 33353 (second panel from top).
This is consistent with an upper bound of $T_\mathrm{i,ped}$ for ELM suppression.
The peripheral density (top panel) and plasma rotation (measured is the
boron, $B^{5+}$, impurity rotation, bottom panel) are identical
at the time of the transition to ELM suppression, hence these quantities
cannot explain the delayed transition in shot 33595.
We will discuss a possible reason for this behaviour
in section \ref{sec:Discussion}.


\subsection{Edge safety factor}
\label{sec:Safety_factor}

\begin{figure*}[t]
  \centering
  \includegraphics[width=0.95\textwidth]{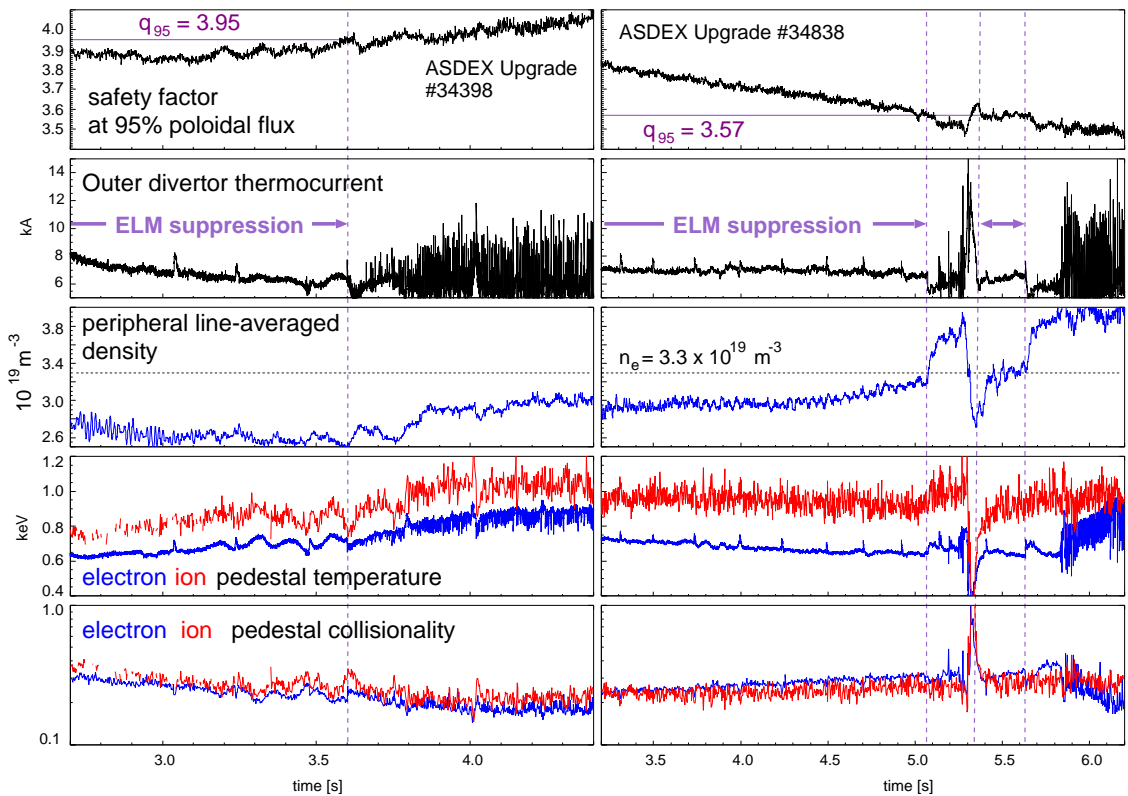}

  \caption{Scans of the edge safety factor $q_{95}$, which is ramped up (left column)
    and down (right column) by slow plasma current variation.
    ELMs remain suppressed in the interval $3.57 < q_{95} < 3.95$.}
  \label{fig:qscans}
\end{figure*}

The existence of safety factor windows for access to ELM suppression has been
reported for DIII-D with $n=3$ \cite{Fenstermacher_PoP15_2008_056122,SCHMITZ_NF52_2012_43005}
and $n=2$ \cite{LANCTOT_NF53_2013_83019} MP.
First experiments are aimed to explore whether similar restrictions exist
in AUG. The safety factor is varied by slow ramps of the plasma current,
with poloidal field coils ramped accordingly to preserve the plasma shape
and plasma volume.
The pulses are started up similarly to the case shown in Fig. \ref{fig:Traces_AUG_33595}
to enter ELM suppression early, followed by the $q_{95}$ ramp until ELM suppression
is lost.
Time traces of two of these discharges are shown in Fig. \ref{fig:qscans},
where transitions to and from ELM suppression are indicated by vertical dashed lines.
In shot 34398, the plasma current is ramped down and ELM suppression is lost
as $q_{95} = 3.95$ is reached. In shot 34838, a lower $q_{95}$ limit is encountered
at $q_{95} = 3.57$.
While ELM suppression is maintained, the peripheral density (third panel from top)
and ion collisionality (bottom panel)
remain below $n_{e,p} = 3.3 \times 10^{19}$~m$^{-3}$ and $\nu^*_\mathrm{i,ped}=0.3$,
respectively, well in the parameter range for ELM suppression.
The loss of ELM suppression is detected as a sharp drop of divertor thermocurrent.
Pedestal parameters change afterwards, in response to the loss of ELM suppression.
We therefore conclude that the $q_{95}$
variation is causal for the back transition and that an access window for
ELM suppression in AUG exists for $q_{95} = 3.57 - 3.95$.
More windows above and below the probed $q_{95}$ range may exist,
but they still need to be explored experimentally.

\begin{figure}[t]
  \centering
  \includegraphics[width=0.7\columnwidth]{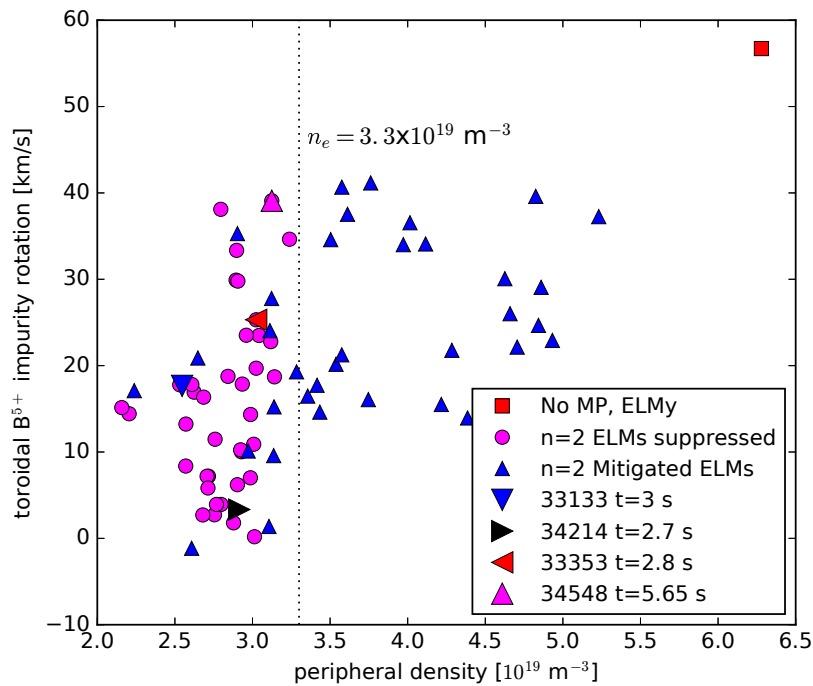}

  \caption{Pedestal impurity ion (B$^{5+}$) rotation velocity (taken at $\psi_n = 0.8$) vs.
    peripheral plasma density. Four individual ELM suppression cases are marked up
    with shot numbers and times of interest - they are used for detailed analysis.}
  \label{fig:vt24_NELP}
\end{figure}

It has been speculated that the reason for the occurrence of $q_{95}$ windows
is the need for resonant surfaces to be placed at certain radial positions near the
pedestal top in order to avoid the expansion of the H-mode edge gradient region
towards destabilisation of ELMs \cite{WADE_NF55_2015_23002}.
From this viewpoint, it is interesting to compare the $q_{95}$ access window
in AUG with those reported for DIII-D.
Width and central $q_{95}$ values for access windows with $n=2$ MP in DIII-D
depend on the differential phase $\Delta \Phi$,
i.e. the relative strength of the plasma response \cite{LANCTOT_NF53_2013_83019}.
For $n=2$ and optimum $\Delta \Phi$, a window centered at $q_{95} = 3.72$ was found
in DIII-D \cite{LANCTOT_NF53_2013_83019},
which can be compared with the center value of $q_{95}=3.76$ in AUG.

It is instructive to also consider the corresponding $n=3$ window documented
for DIII-D \cite{SCHMITZ_NF52_2012_43005}, $q_{95} = 3.77-3.91$.
Because of the different fractional resonant surfaces
(AUG: $q=m/2$, DIII-D: $q=m/3$, where $m$: integer),
the similarity of the upper $q_{95}$ limit
in both machines suggests that it corresponds to the position of an integer and
not a non-integer rational surface.
The integer surface next to the top of the gradient region is, in both cases,
the $q=4$ surface. The next lower resonant surface
($q=7/2$ in AUG, $q=11/3$ in DIII-D)
will take this same position at $q_{95} \sim 3.42$ in AUG and $q_{95} \sim 3.67$ in DIII-D,
which should therefore represent the upper $q_{95}$ bound of the next
ELM suppression access window. 
For DIII-D, this matches the experimental value of $q_{95}=3.65$ reported
in Ref. \cite{SCHMITZ_NF52_2012_43005}.
For AUG, there is no such reference as no safety factor scan at lower $q_{95}$
has been made to date.
A more direct comparison would be comparing
different $n$ values in the same machine, however, ELM suppression has not been observed
with $n=3$ MP in AUG to date.


\subsection{Plasma rotation}
\label{sec:Rotation}

\begin{figure}[t]
  \centering
  \includegraphics[width=0.7\columnwidth]{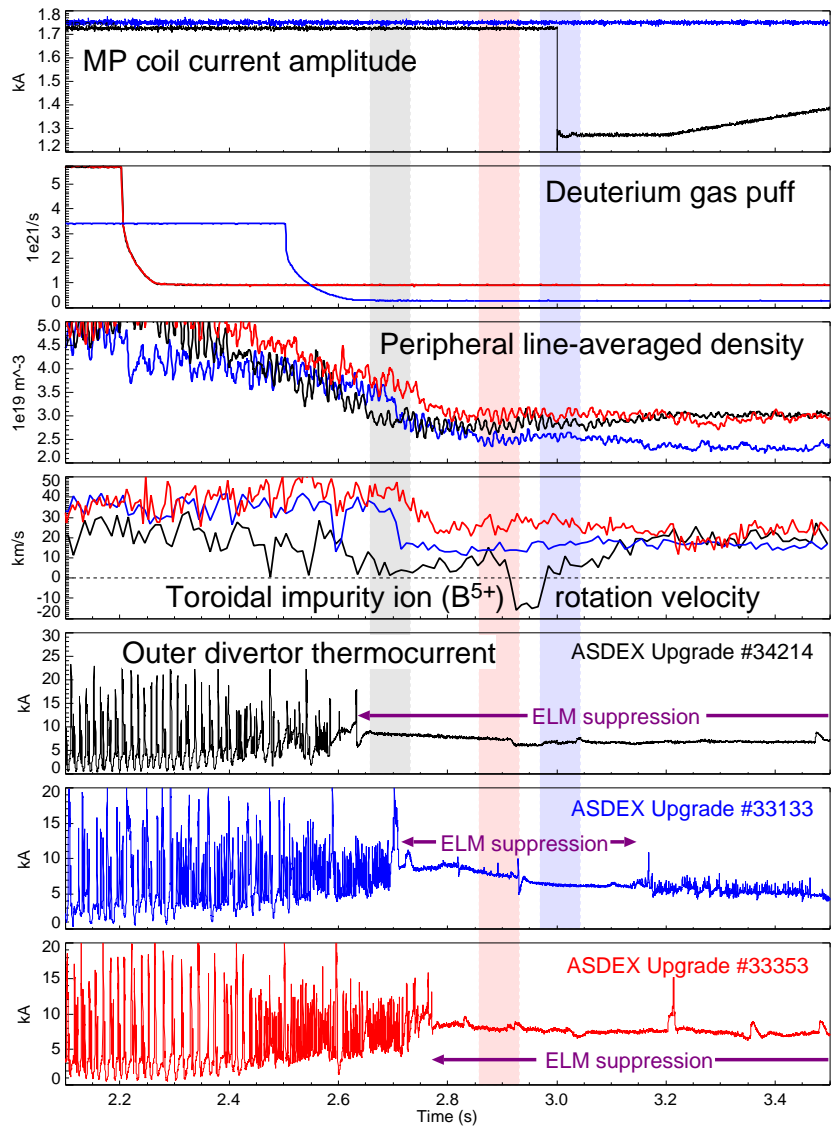}

  \caption{Time traces of discharges 33133, 33353, 34214 around
    the transition to ELM suppression, showing variations of pedestal
    toroidal impurity ion rotation.}
  \label{fig:Traces_34214_33133_33353}
\end{figure}

A recent study \cite{MOYER_PP24_2017_102501} showed that access to ELM suppression
in DIII-D depends on the torque applied to the plasma by neutral beam injection,
leading to a threshold in plasma rotation.
For small flows or flows directed in counter-current direction,
ELM suppression could not be obtained.
Depending on the underlying physics reason, this is a potential issue for ITER
and a fusion reactor where small plasma rotation is expected in the absence
of strong external momentum sources.
Significant variation of plasma rotation is encountered in ELM suppression
discharges in our present experiment.
Fig. \ref{fig:vt24_NELP} shows the toroidal rotation velocity
of boron ($B^{5+}$) impurities, measured by a charge exchange recombination spectroscopy
(CXRS) sightline which intersects one of the heating neutral beams at
normalised poloidal flux $\psi_n = 0.8$, i.e. on the pedestal top, for the
plasma shape used in these experiments.
The data set of Fig. \ref{fig:nelp_nuped} is used, with the
same symbol and colour coding, but without restrictions for $\Delta \Phi$
in order to represent our full set of ELM suppression cases.
Again, only time intervals stationary for at least $100$~ms are shown.
One can see that ELM suppression is observed in a large range of impurity velocities,
v$^{B5+}_\mathrm{tor} = 0-40$~km/s and that no separation in rotation velocity
between ELM suppression and ELM mitigation is visible in the
toroidal rotation velocity range covered in our experiments so far.

In Fig. \ref{fig:vt24_NELP}, four ELM suppression cases
(triangles with different orientations and colours)
are marked up with their shot numbers and times of interest.
The toroidal impurity rotation for these cases is different, and
we will study them in more detail subsequently.
Time traces for three of these four cases
are shown in Fig. \ref{fig:Traces_34214_33133_33353}.
The full duration of the ELM suppressed state in each discharge is
indicated by horizontal arrows labelled ``ELM suppression''.
The transition to ELM suppression occurs at different values of the toroidal
impurity rotation and is dictated by the time the plasma density drops
below $n_\mathrm{e,p}=3.3 \times 10^{19}$~m$^{-3}$.
However, the plasma rotation drops somewhat after this transition in shots
$33133$ and $33353$ where it was initially high, indicating a stronger
braking torque during ELM suppression than during ELM mitigation.

\begin{figure*}[t]
  \centering
  \includegraphics[width=0.8\textwidth]{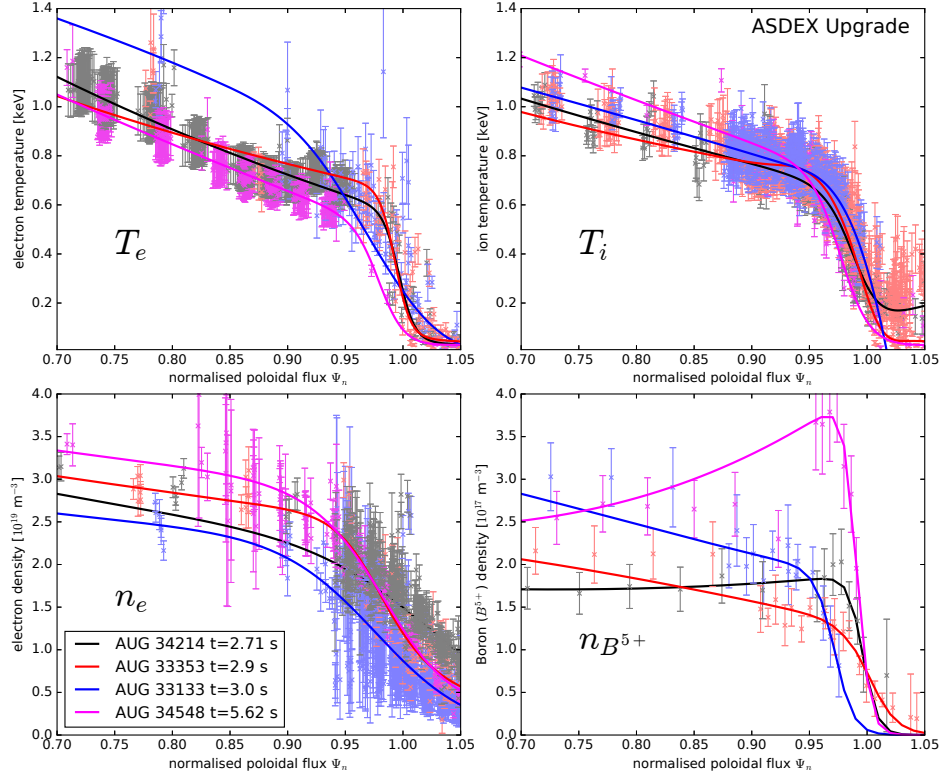}

  \caption{Profiles of $T_e$, $T_i$, $n_e$ and $n_{B^{5+}}$
    (experimental data points with error bars
    and smooth fitting curves as solid lines) in the edge pedestal region
    for discharges 34214, 33133, 33353, 34548 at the time points indicated.}
  \label{fig:ped_profiles}
\end{figure*}

For further analysis, we pick one time interval in each of these discharges
during fully established ELM suppression.
They are indicated by vertical shaded areas in Fig. \ref{fig:Traces_34214_33133_33353},
which are coloured similarly to the measured curves of each discharge.
A fourth time interval is taken during a long stationary ELM suppression phase
in pulse $34548$ around $t=5.65$~s.
Fig. \ref{fig:ped_profiles} shows profiles of $T_e$, $T_i$, $n_e$ and
$n_{B^{5+}}$ (density of fully stripped boron impurity ions) in the edge
pedestal region, originating from core and edge Thomson scattering ($T_e$, $n_e$),
core and edge CXRS on boron impurities ($T_i$, $n_{B^{5+}}$), and Li~beam ($n_e$).
Hyperbolic tangent fits to this data are shown as solid lines.
Fits to the density are constrained by the DCN interferometer line integrals
in addition to radially resolved profiles.
The edge gradient and the pedestal top regions are well resolved by these measurements
so that electron and ion diamagnetic velocity profiles can be determined.
There is little variation of $T_e$, $T_i$ and gradients of these quantities.
Boron is the prevalent light impurity species and occurs with about 1\% or lower concentration.
The impurity density shows a clear pedestal and a steep gradient at $\psi_n>0.95$.

\begin{figure}[t]
  \centering
  \includegraphics[width=0.7\columnwidth]{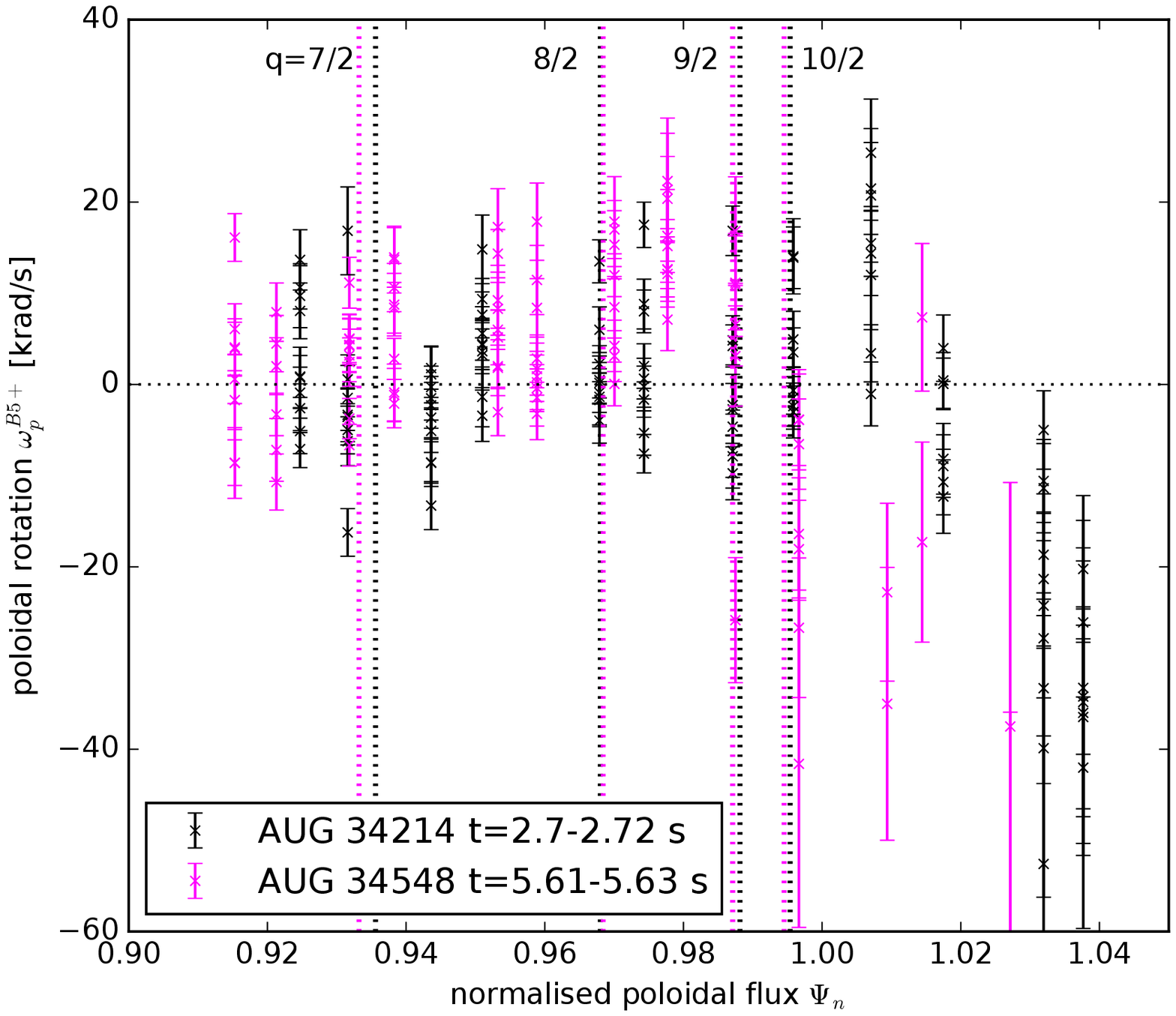}

  \caption{Poloidal rotation profile
    in the pedestal region at the outer midplane for two time intervals,
    pulse 34214, $t=2.7-2.72$~s and 34548, $t=5.61-2.5.63$~s.
    Nominal positions of resonant surfaces are marked by vertical dashed lines}
  \label{fig:vpol_34214}
\end{figure}

We will now examine these four cases in view of a recent model
for ELM suppression \cite{WADE_NF55_2015_23002} which invokes
an unshielded resonant response to the MP to block the expansion of the
edge transport barrier before an ELM crash can occur. 
In various linear two-fluid MHD calculations
\cite{YU_NF51_2011_073030,FERRARO_PP19_2012_056105},
the cross-field electron flow v$_{e,\perp}$ has been identified to
control the shielding of the external MP at rational surfaces.

In AUG, impurity ion flows are measured in toroidal (v$_{\alpha,t}$)
and poloidal (v$_{\alpha,p}$) directions by charge exchange recombination
spectroscopy \cite{VIEZZER_RSI83_103501}.
The index $\alpha$ denotes the impurity species used,
fully stripped boron ($B^5+$) with charge state $Z_\alpha = 5$ 
in the present experiment.
We obtain v$_{e,\perp}$ from the combined radial force balances
of electrons and impurity ions \cite{VIEZZER_NF53_2013_053005},
\begin{equation}
  \mathrm{v}_{e,\perp}
  = \frac{\nabla p_e}{e n_e B} + \frac{E_r}{B} 
  = \frac{\nabla p_e}{e n_e B} + \frac{\nabla p_\mathrm{\alpha}}{Z_\mathrm{\alpha} e n_\mathrm{\alpha} B}
  + \mathrm{v}_\mathrm{\alpha,t} \frac{B_p}{B} - \mathrm{v}_\mathrm{\alpha,p} \frac{B_t}{B}
  \label{Eq:ForceBalance}
\end{equation}
where $e$ is the elementary charge; $B_t$, $B_p$, and $B=(B_t^2 + B_p^2)^{1/2}$
are the toroidal, poloidal
and total magnetic inductance, respectively.
$\nabla p_e / (e n_e B)$ and
$\nabla p_\mathrm{\alpha} / (Z_\mathrm{\alpha} e n_\mathrm{\alpha} B)$
are the electron and impurity diamagnetic flows, respectively,
and the last two terms represent the cross field impurity flow.
Often the terms of the force balance are expressed as angular frequencies $\omega$,
with the advantage that most of them become flux functions and can more easily
be compared with numerical code output. Eq. \ref{Eq:ForceBalance} then becomes
\begin{equation}
  \omega_{e,\perp}
  = \frac{p^\prime_e}{e n_e} + \frac{E_r}{|R B_p|}
  = \frac{p^\prime_e}{e n_e} + \frac{p^\prime_\mathrm{\alpha}}{Z_\mathrm{\alpha} e n_\mathrm{\alpha}}
  +  \frac{\mathrm{v}_\mathrm{\alpha,t}}{R} - \mathrm{v}_\mathrm{\alpha,p} \frac{B_t}{|R B_p|}
  \label{Eq:ForceBalance2}
\end{equation}
where now the derivative $p^\prime = \mathrm{d}p / \mathrm{d}\psi$ is with respect to
the poloidal flux ($\psi$ in Vs/rad).
Here, $\omega^*_e = p^\prime_e / (e n_e)$, $\omega_{E \times B} = E_r / (|R B_p|)$,
$\omega^*_\alpha = p^\prime_\mathrm{\alpha} / (Z_\mathrm{\alpha} e n_\mathrm{\alpha})$
and $\omega_{\alpha,\perp} = \omega_{\alpha,t} + \omega_{\alpha,p}$ are flux functions,
while $\omega_{\alpha,t} = (\mathrm{v}_\mathrm{\alpha,t}/R)$
and $\omega_{\alpha,p} = - \mathrm{v}_\mathrm{\alpha,p} (B / |R B_p|)$
are not flux functions individually.

Several observations can be made in the course of the analysis.
In Eq. \ref{Eq:ForceBalance2}, 
the poloidal impurity flow v$_\mathrm{\alpha,p}$ is weighted stronger
by a factor $B_\mathrm{t}/B_\mathrm{p}$ than v$_\mathrm{\alpha,t}$ and hence the errors of
$\omega_\mathrm{\alpha,p}$ typically dominate the errors in the cross field impurity flow.
Fig. \ref{fig:vpol_34214} shows the measured impurity poloidal rotation
$\omega^{B5+}_p$ for two cases,
pulse 34214 in the time interval $t=2.7-2.72$~s
and pulse 34548 at $t=5.61-5.63$~s.
All measurements in these time intervals are overlayed and show significant scatter.
On the pedestal top, the averaged poloidal rotation essentially vanishes within the scatter
of the data and can become significant only
in the gradient region and scrape-off-layer, $\psi_n \geq 0.98$.
Neglecting $\omega^{B5+}_p$ in Eq. \ref{Eq:ForceBalance2} greatly reduces the uncertainty
of the result, and constitutes an upper bound of $\omega_{e,\perp}$.

\begin{figure}[t]
  \centering
  \includegraphics[width=0.7\columnwidth]{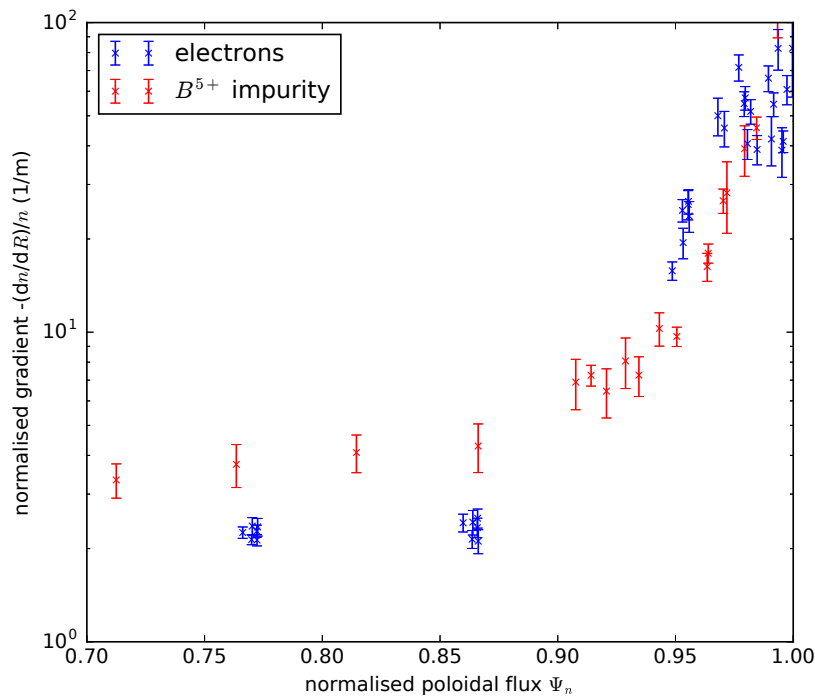}

  \caption{Comparison of $\nabla n / n$ profiles for impurity ions (red) and electrons (blue) 
    at the outer midplane of discharge 33353, $t=2.9$~s.}
  \label{fig:gradn_n}
\end{figure}


The impurity diamagnetic term in Eq.~\ref{Eq:ForceBalance} can be written as
$\nabla p_\mathrm{\alpha} / (Z_\mathrm{\alpha} e n_\mathrm{\alpha} B) = $
$[T_\alpha(\nabla n_\alpha / n_\alpha) + \nabla T_\alpha] / (Z_\alpha e B)$, i.e.
there is no dependence on the absolute impurity density, but only on the
impurity density gradient length $n_\alpha / \nabla n_\alpha$.
Furthermore, for similar density gradient length and temperature of impurity ions
and electrons, the impurity diamagnetic flow is smaller 
by a factor of $Z_\alpha = 5$ (for boron as in our case)
than the electron diamagnetic flow.
The density gradient length for impurity ions and electrons is
often very similar at the pedestal top.
Fig. \ref{fig:gradn_n} shows the case with the most
peaked impurity density profile at the pedestal top among our set of four
highlighted discharges, pulse 33353 at $t=2.9$~s.
The edge transport barrier at $\psi_n > 0.95$ is well seen in both species.
At the pedestal top, the contribution of the density gradient length term
in the impurity ion diamagnetic flow to Eq. \ref{Eq:ForceBalance}
is about $(\nabla n/n = -2$/m$)\times (Ti=800$~eV$)/(Z=5)/(B=1.4$T$)$~$=230$~m/s,
which is small compared to the electron diamagnetic flow.
The accuracy of v$_{e,\perp}$ is therefore mainly determined by the errors
of the electron diamagnetic and the impurity cross-field flows.

\begin{figure}[t]
  \centering
  \includegraphics[width=0.7\columnwidth]{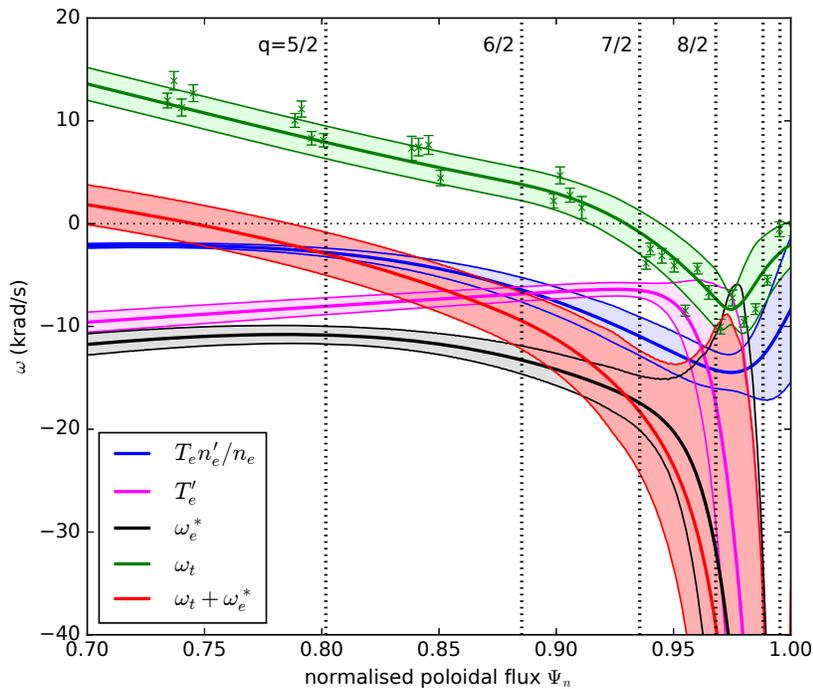}

  \caption{Profiles of angular rotation frequency of various terms in the
    force balance Eq. \ref{Eq:ForceBalance2}:
    The electron diamagnetic velocity $\omega^*_e$ (black) and its components
    $T_e (n_e^\prime / n_e)$ (blue) and $T_e^\prime$ (magenta),
    the toroidal impurity flow $\omega_t$ (green),
    and the sum of $\omega^*_e$ and $\omega_t$ (red)
    in the edge pedestal region
    for discharge 34214, $t=2.71$~s.}
  \label{fig:omega_err}
\end{figure}

We inspect now the dominant terms in the radial force balance, Eq. \ref{Eq:ForceBalance2},
for one example, pulse 34214 at $t=2.71$~s.
Fig. \ref{fig:omega_err} shows angular frequencies of the electron diamagnetic rotation
$\omega^*_e$ (black curve), its components $T_e n_e^\prime / (e n_e)$ (blue curve)
and $T_e^\prime / e$ (magenta curve), the toroidal rotation $\omega_t$ (green curve)
along with the original measurement (green symbols)
and the sum of $\omega_t$ and $\omega^*_e$ (red curve).
Least squares fits to the original diagnostic data are applied in order to calculate
the rotation angular frequencies on a common dense grid of $\psi_N$.
The coloured bands represent propagated experimental errors, profile fit errors
and errors of the radial alignment between the various diagnostics.
While $\omega_t$ (green) changes sign near the pedestal top, and
in this case remains small in the entire pedestal region,
$\omega^*_e$ is strictly in the electron diamagnetic (negative) direction.
Their sum $\omega_t$ and $\omega^*_e$ corresponds to $\omega_{e,\perp}$ as given 
by the force balance Eq. \ref{Eq:ForceBalance2}, but without $\omega_{\alpha,p}$
(small or negative, Fig. \ref{fig:vpol_34214}) and without $\omega^*_{\alpha}$ (small).
In this example, $\omega_t + \omega^*_e$ (red) crosses zero at $\psi_n \approx 0.75$
and is negative (outside error bars) for $\psi_n > 0.8$, i.e. in the entire pedestal
top and gradient regions. This includes the locations of the $q=8/2$ and $q=7/2$
surfaces, which are near the upper end of the gradient region and therefore are
candidates for a resistive plasma response to the MP.

\begin{figure*}[t]
  \centering
  \includegraphics[width=\textwidth]{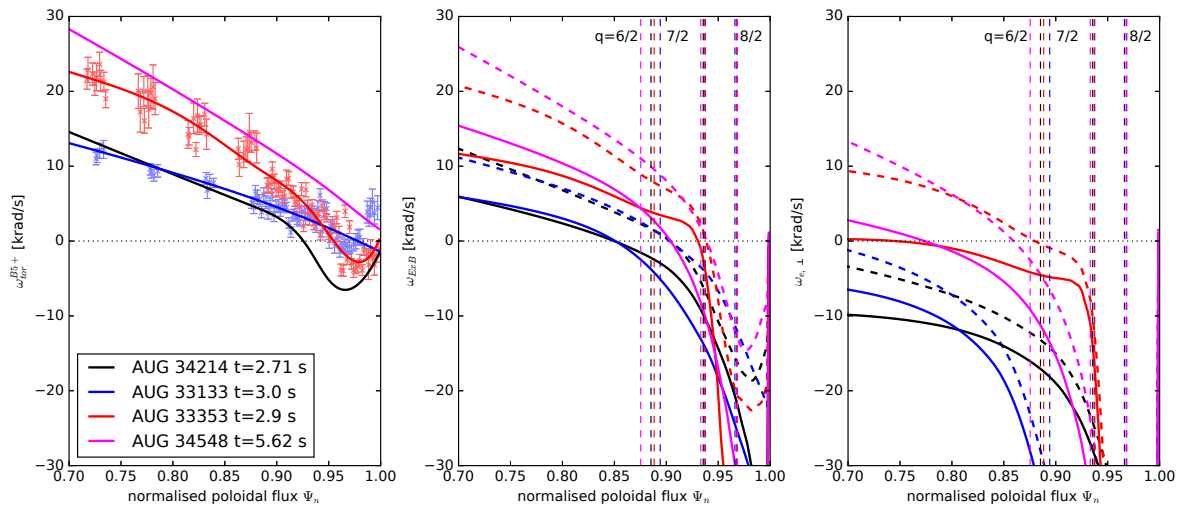}

  \caption{Profiles of angular rotation frequency of impurity ions ($B^{5+}$, left panel),
    gyrocentres ($E \times B$ flow, middle panel) and electron fluid perpendicular
    to $B$ ($\omega_{e,\perp}$, right panel) in the edge pedestal region
    for discharges 34214, 33133, 33353, 34548 at the time points indicated.
    Solid curves are calculated with neoclassical $\omega^{B5+}_p$, dashed curves
    with $\omega^{B5+}_p = 0$.
    The position of various resonant surfaces is marked by vertical dashed lines.}
  \label{fig:rot_profiles}
\end{figure*}

For our four cases of interest, we now evaluate the full force balance,
Eq.~\ref{Eq:ForceBalance2}, including $\omega^*_{\alpha}$ and $\omega_{\alpha,p}$.
In order to avoid the errors associated with the $\omega_{\alpha,p}$ measurement,
we use the neoclassical estimate for $\omega_{\alpha,p}$ from the 
NEOART code \cite{DUX99A,PEETERS00A}.
In a previous study of H-mode plasmas in AUG \cite{VIEZZER_NF55_2015_123002},
which included low H-mode pedestal collisionalities ($\nu^*_\mathrm{i,ped} \ll 1$),
good agreement was found between measured and neoclassical poloidal rotation.
For our present discharges we find that the neoclassical calculation
tends to slightly underestimate $\omega^{B5+}_{p}$
(predict more negative values than measured) on the pedestal top.
Fig.~\ref{fig:rot_profiles} shows
the measured pedestal rotation profiles of the impurities ($B^{5+}$) in toroidal direction
$\omega_\mathrm{t} = \mathrm{v}_\mathrm{t} / R$ (left panel, with experimental errors),
the gyrocentres $\omega_{E\times B} = E_r / |R B_p |$ (middle panel)
and the cross field flow of the electron fluid
$\omega_{e,\perp} = \omega_{E\times B} + p^\prime_e / (e n_e)$ (right panel).
Solid curves in the middle and right panel represent the values obtained using the
full force balance, Eq.~\ref{Eq:ForceBalance2}, including neoclassical $\omega^{B5+}_{p}$.
Dashed curves are calculations with $\omega^{B5+}_{p}$ assumed to be zero,
which represents an upper bound of $\omega_{E \times B}$ and $\omega_{e,\perp}$,
as discussed above.
The $E \times B$ rotation (middle panel) changes sign at the plasma edge in all
our cases, because with co-injected neutral beams as used in all our present discharges,
$\omega_{E \times B} > 0$ (ion diamagnetic direction) in the core, while 
in the edge gradient region ($\psi_n > 0.93$), poloidal and diamagnetic flows
always drive a strong inward directed radial electric field, $\omega_{E \times B} < 0$.
The precise position of $\omega_{E \times B} = 0$ depends crucially on the
actual errors of the analysis, in particular the precision of $\omega^{B5+}_{p}$.
At the present time we cannot determine whether or not $\omega_{E \times B} = 0$
is aligned with rational surfaces or not.

Because of a significant electron diamagnetic rotation $\omega_\mathrm{e}^*$, 
$\omega_{e,\perp}$ is clearly offset from $\omega_{E \times B}$.
As shown in the right panel of Fig.~\ref{fig:rot_profiles},
the electron perpendicular rotation has zero crossings $\omega_{e, \perp} = 0$
for two of our four selected cases and no zero crossings for the other two,
independent of whether $\omega^{B5+}_{p}$ is neglected or taken
from the neoclassical calculation. Again, it should be noted
that for our present discharges this choice corresponds approximately to an upper or lower
bound for the true value of $\omega_{e, \perp}$, respectively.
At the $q=7/2$ and $q=8/2$ resonant surfaces, i.e. near the inner end of the edge gradient region,
$| \omega_{e,\perp} |$ becomes large for all our cases.
We compare this result with shielding calculations and discuss the implications
of our findings in the next section (section \ref{sec:Discussion}).


\section{Summary and Discussion}
\label{sec:Discussion}

In many respects, the ELM suppression regime
in ASDEX Upgrade at low pedestal collisionality
resembles that of the original DIII-D discovery:
ELMs are suppressed after a sharp transition encountered normally from
phases with ELMs, which are typically mitigated already by the MP.
The mode number spectrum of the MP in both machines matters in
that optimum coupling to amplifying edge pedestal-driven kink-peeling
modes is essential for ELM suppression access.
During ELM suppression phases, significant particle transport across the H-mode
edge transport barrier occurs, and the plasma density with identical fuelling
is usually below that of ELMy phases, despite the absence of ELMs
and ELM-related particle losses.
Plasma density and stored energy are stationary for many confinement times,
if the MP is continuously applied and sufficiently strong to keep
the plasma density below a limit which is very similar in AUG and \mbox{DIII-D}.
This upper density limit for ELM suppression can be expressed
by a maximum value near $n_\mathrm{e,ped} = 3 \times 10^{19}$~m$^{-3}$
of pedestal top plasma density or as
a maximum pedestal collisionality near $\nu^*_\mathrm{i,ped} = 0.3$.
Since AUG and \mbox{DIII-D} have about the same physical size, it is not possible
to identify which density-related dimensionless parameter describes
the actual physical requirement for achieving ELM suppression.
Finally, within the range of edge safety factor $q_{95}$ examined so far in AUG,
one $q_{95}$ window for ELM suppression has been detected that seems to have
a clear corresponding $q_{95}$ window in DIII-D, despite different plasma
shapes.
The $q_{95}$ access window width for our experiment with $n=2$ MP is wider
than the corresponding window's width of DIII-D ($n=3$), as expected for 
the sparser radial distribution of resonant surfaces
(half integer instead of third integer $q=m/n$).
These observations are consistent with the assumption that the location of
resonant surfaces and therefore a resistive response play an important
role for ELM suppression.

However, there is an apparent insensitivity to plasma rotation variations
and therefore, varying conditions for shielding of a resistive response.
We observe ELM suppression in cases where the pedestal top
impurity rotation is very small as expected for a burning
plasma without external momentum input, and consequently
the electron cross-field flow $| \omega_{e,\perp} |$ is large.
In the DIII-D experiment \cite{MOYER_PP24_2017_102501},
input torque variations around zero net torque have been produced by a mixture of
co-$I_p$ and counter-$I_p$ NBI, which is not possible in AUG.
All our plasma have been heated with co-$I_p$ directed NBI and the
variation of plasma rotation originates mainly from plasma density
and MP field strength variations.
Despite this technical limitation, impurity rotation varies
widely in AUG, v$^{B5+}_\mathrm{tor} = 0-40$~km/s, and
as shown in section \ref{sec:Rotation},
there is a concomitant strong variation of $\omega_{e,\perp}$.

This rotation variation can be compared with the cross-field electron flow
required for shielding the resistive response in linear MHD model predictions.
For ELM suppression plasmas in AUG and DIII-D several such calculations have
been made \cite{LIU_NF56_2016_56035,LIU_POP24_2017_056111,LYONS_PPCF59_2017_044001}.
Single-fluid \mbox{MARS-F} calculations for the AUG experimental case
\cite{LIU_POP24_2017_056111} show that a fairly
small cross-field flow, of the order of $|\omega| \leq 6$~krad/s, is required
to obtain a significant resistive response at a resonant surface.
A similar study has been carried out for DIII-D equilibria, using a two-fluid MHD model
implemented in the M3D-C1 code \cite{LYONS_PPCF59_2017_044001}.
This study shows that the resonant response for a single row of MP coils in \mbox{DIII-D}
as a function of electron cross-field rotation is strongly peaked,
with a half width of $| \omega_{e,\perp} | \leq 5$~krad/s
around maximum response (section 3.1 in Ref. \cite{LYONS_PPCF59_2017_044001}).
In this respect, this result agrees with that of Ref. \cite{LIU_POP24_2017_056111}.
However, the maximum response is found to not coincide exactly with zero flow
at the resonant surface location, but is slightly skewed in radius to either side
of the resonance, depending on whether the upper or lower MP coil ring is
considered.
The authors of Ref. \cite{LYONS_PPCF59_2017_044001}
do not give an explanation for this effect in their modelling.
We do not have the same two-fluid analysis for AUG, but we can inspect
our experimental data presented in section \ref{sec:Rotation}
whether the electron cross-field flow is small,
$| \omega_{e,\perp} | \leq 5$~krad/s,
in the vicinity of resonant surfaces in the edge pedestal region,
even if not exactly aligned with a surface.
This is true for none of the cases of Fig. \ref{fig:rot_profiles} 
at the $q=8/2$ surface,
and for shots 34214 and 33133 there is no region at the pedestal for which
$| \omega_{e,\perp} | \leq 5$~krad/s.

If a resistive response is important for ELM suppression at all, 
it is difficult to understand our observations from the viewpoint
of a linear MHD description of the plasma response.
Kinetic modelling \cite{HEYN_NF54_2014_064005} suggests that 
guiding center orbit resonances at $\omega_{E \times B} = 0$
(for stationary or slowly varying MP)
play a role for field penetration and particle transport.
In our present experiments, a surface with $\omega_{E \times B} = 0$ exists
because of co-current (positive) $E \times B$ rotation in the core and the
inward directed $E_r$ well, i.e. negative $\omega_{E \times B}$, in the H-mode barrier.
Consequently, $\omega_{E \times B} = 0$ in the vicinity of the inner boundary
of the gradient region.
It is a remaining task to develop and apply kinetic models to
AUG ELM suppression experiments
and explore the sensitivity of ELM suppression to the $\omega_{E \times B} = 0$
location.

\begin{figure}[t]
  \centering
  \includegraphics[width=0.7\columnwidth]{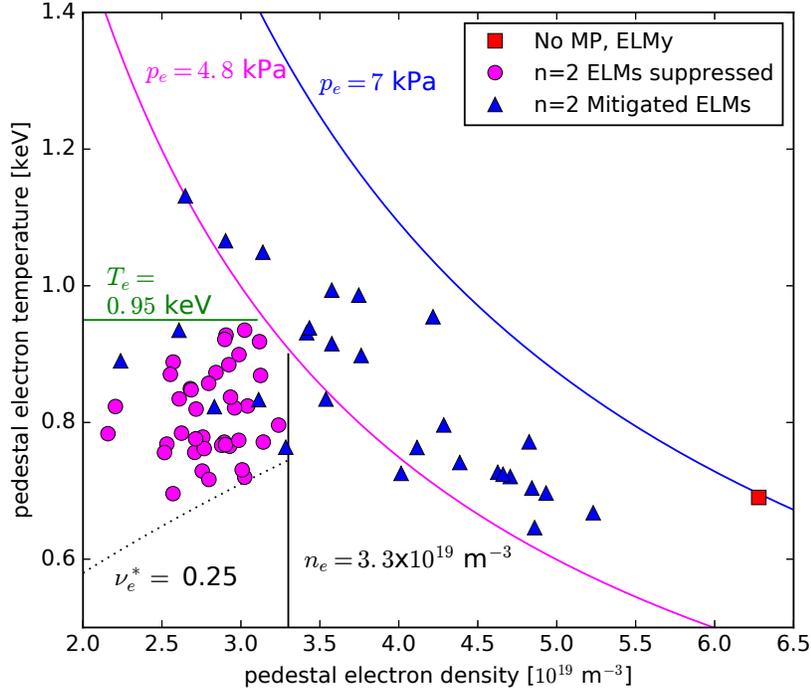} 
  \caption{Operational boundaries in pedestal $T_e$-$n_e$ space
    of ELM suppression (circles), ELMy H-mode with
    MP-mitigated small ELMs (triangles), and ELMy H-mode with MP off (red square).}
  \label{edgeopdia}
\end{figure}

A surprising finding in AUG is the lack of ELM suppression
at ITER-relevant low edge pedestal collisionality $\nu^*_\mathrm{i,ped} \leq 0.15$,
despite sufficiently low density for ELM suppression.
This can be attributed to a high pedestal temperature
(section \ref{sec:EdgeDensCollisionality}).
Another view emerges if one examines the locus of ELM suppression
and ELM mitigation in edge pedestal temperature-density space,
also referred to as the H-mode edge operational diagram \cite{SUTTROP97C}.
Fig.~\ref{edgeopdia} shows electron parameters, $T_\mathrm{e,ped}$ vs. $n_\mathrm{e,p}$,
for the AUG ELM suppression data set
together with an annotation of empirical regime boundaries.
Only cases with $q_{95}=3.57 \ldots 3.95$, i.e. within the safety factor
access window, and with the same nominal plasma shape are selected. 
The cases of returning small ELMs at low collisionality ($\nu^*_\mathrm{i,ped} \leq 0.15$)
appear above a temperature threshold, $T_e \geq 1.0$~keV (green line).
They are also close to a line of constant pedestal electron pressure (magenta line)
at $p_e = 4.8$~kPa which is bounding the actual ELM suppression cases, and which
is decorated by most ELM mitigation cases at higher density and lower temperature.
We can therefore hypothesise that the return of ELMs at low collisionality
is due to the pedestal reaching a stability limit for
triggering small ELMs with applied MP.
This stability limit is considerably reduced compared to ELMy H-mode without MP.
As a reference without MP, our case with lowest edge density (red square)
has considerably larger edge pressure, $p_e = 7$~Pa (blue curve).
Therefore, and in addition to the density reduction by the ``pump-out'' effect,
a reduction of pedestal pressure appears as an additional price for
ELM mitigation or ELM suppression despite access
to higher pedestal temperature at low density.
As H-mode confinement depends largely on pedestal properties,
it is of high interest for the fusion performance of ITER and future fusion devices
to examine the reason for the observed pedestal pressure reduction and devise
ways to minimise it.

A possible reason for the reduced edge stability with MP applied has been
pointed out in a recent study of toroidally localised inter-ELM oscillations in AUG
with applied MP \cite{WILLENSDORFER_PRL119_85002}.
The MP causes toroidal variations of the local magnetic shear which 
destabilise ballooning modes in a toroidally restricted region, for field lines where,
experimentally, the inter-ELM oscillation is observed. 
It can therefore be expected that the maximum stable edge pressure gradient is
reduced when the MP is applied.
The situation is complicated by the fact that for low collisionalities,
such as in our cases near ELM suppression,
a strong bootstrap current exists in the gradient region,
which leads to destabilisation of medium-$n$ edge peeling modes
that couple with infinite-$n$ ballooning modes \cite{WILSON_POP6_1999_1925}.
Linear \cite{LIU_POP24_2017_056111}
and non-linear \cite{ORAIN_PP20_2013_102510, STRUMBERGER_NF54_2014_64019}
MHD models have so far been mainly used to predict the plasma response
to the applied low-$n$ MP, with quantitative success to describe the plasma edge displacement
in AUG \cite{WILLENSDORFER_NF57_2017_116047}.
Wingen {\em et al} \cite{Wingen_PPCF57_2015_104006} find for selected DIII-D cases
that at low pedestal collisionality the increased \mbox{H-mode} edge bootstrap current 
leads to both larger helical plasma deformation
and stronger destabilisation of peeling-ballooning modes than at high collisionality.
That would suggest a reduced stability limit at high edge temperature and possibly
explain the re-appearance of small ELMs.
This question can be addressed in the future by edge stability calculations
for a 3D equilibrium against a wide range of modes, such as coupled peeling-ballooning modes, 
and quantitative comparisons with empirical edge stability limits in AUG.

From Fig. \ref{edgeopdia}, we note that, with the exception of a few cases of
mitigated ELMs at very low pressure, all ELM suppression cases seem to be grouped
below the pedestal pressure associated with mitigated ELMs.
This suggests that lifting the small ELM pressure gradient may lead to an extension
of the edge operational range for ELM suppression access.
Edge stability can be improved by stronger shaping of the plasma cross-section
which allows to maintain larger pressure gradient and pedestal pressure
without triggering ELMs. As an additional benefit, increased pressure gradient and
bootstrap current may increase the drive for amplification of the externally applied MP
by marginally stable low-$n$ peeling modes, and hence reduce the MP coil current threshold
for ELM suppression.
We can speculate that the required increased triangularity in AUG to achieve ELM suppression
\cite{NAZIKIAN_FEC2016, SUTTROP_PPCF59_014049} is caused by a combination
of these two factors.

A few observations in Fig. \ref{edgeopdia} remain unexplained so far.
The existence of an upper density (black solid curve) or collisionality (black dotted curve)
limit (shown is the locus of $\nu^*_e=0.25$ which bounds our data)
cannot be explained solely by a pressure-driven stability argument.
The small temperature variation near this boundary in our present data also does not
allow us to distinguish conclusively between these two parameters
(or a possible third, density-related, parameter).
Variation of $Z_\mathrm{eff}$ by seeding with low-$Z$ impurities such as nitrogen,
and variation of the major radius $R$, i.e. comparison of plasmas in machines
with different size, would probably be most effective to test a collisionality boundary.
The other observation is the re-appearance of very small ELMs at low edge pedestal pressure
in a few cases (blue triangles well below the magenta line in Fig. \ref{edgeopdia}), which
can take the form of sharp, seemingly unmotivated, transitions out of suppression.
So far no parameter has been identified in our data set that triggers these transitions.
This question requires more attention in upcoming experiments in AUG.


\section*{Acknowledgments}
W.S. and E.V. thank R. Dux for valuable discussions.
This work has been carried out within the framework of the 
EUROfusion Consortium and has received funding from the 
Euratom research and training programme 2014-2018 
under grant agreement No 633053. 
The views and opinions expressed herein do not necessarily 
reflect those of the European Commission.

\section*{References}

\bibliographystyle{unsrt}
\bibliography{elmsuppacc}

\end{document}